\documentclass[twocolumn,showpacs,prl]{revtex4}

\usepackage{graphicx}
\usepackage{dcolumn}
\usepackage{bm}
\usepackage{color}

\begin{document}

\preprint{???}

\title{Nanosecond molecular relaxations in lipid bilayers studied by high energy resolution neutron scattering and in-situ diffraction}

\author{Maikel~C.~Rheinst\"adter$^1$}\email{rheinstaedter@ill.fr}
\author{Tilo Seydel$^1$}
\author{Tim~Salditt$^2$}

\affiliation{$^1$Institut Laue-Langevin, 6 rue Jules Horowitz, BP
156, 38042 Grenoble Cedex 9, France\\
$^2$Institut f\"{u}r R\"{o}ntgenphysik, Friedrich-Hund Platz 1,
37077 G\"{o}ttingen, Germany}

\date{\today}

\begin{abstract}
We report a high energy-resolution neutron backscattering study to
investigate slow motions on nanosecond time scales in highly
oriented solid supported phospholipid bilayers of the model system
DMPC -d54 (deuterated
1,2-dimyristoyl-sn-glycero-3-phoshatidylcholine), hydrated with
heavy water. Wave vector resolved quasi-elastic neutron scattering
(QENS) is used to determine relaxation times $\tau$, which can be
associated with different molecular components, i.e., the lipid
acyl chains and the interstitial water molecules in the different
phases of the model membrane system. The inelastic data are
complemented both by energy resolved and energy integrated in-situ
diffraction. From a combined analysis of the inelastic data in the
energy and time domain, the respective character of the
relaxation, i.e., the exponent of the exponential decay is also
determined. From this analysis we quantify two relaxation
processes. We associate the fast relaxation with translational
diffusion of lipid and water molecules while the slow process
likely stems from collective dynamics.
\end{abstract}

\pacs{87.14.Cc, 87.16.Dg, 83.85.Hf, 83.10.Mj}

\maketitle

\section{Introduction\label{Introduction}}
\begin{figure*}[] \centering
\resizebox{0.5\textwidth}{!}{\rotatebox{0}{\includegraphics{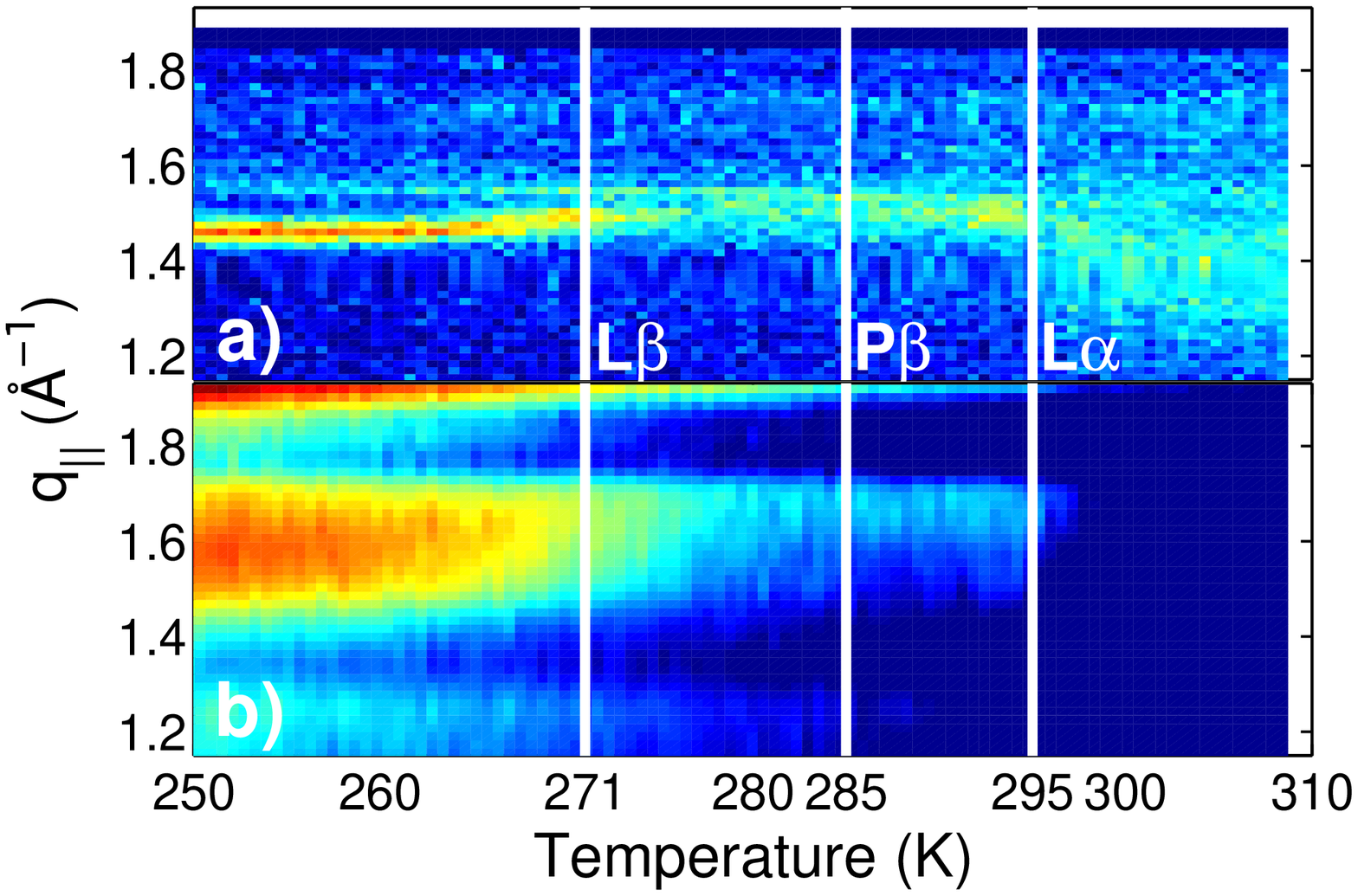}}}
\begin{minipage}[b]{8.85cm}
\centering \caption[]{(Color online). (a) Energy integrated
diffraction for temperatures 250~K$<$T$<$310~K with a temperature
resolution of $\Delta T\approx$0.3~K (normalized to acquisition
time). $q_{||}$ is the in-plane component of the scattering vector
$\vec{Q}$. The phase boundaries for gel (L$\beta$), ripple
(P$\beta'$) and fluid phase (L$\alpha$) as defined by structural
changes, i.e., the change of peak position and width are marked by
the solid white lines. (b) 'True' elastic scattering, measured
with an energy resolution of 0.9~$\mu$eV. The phase transitions
are also visible in the energy resolved diffraction data by
following the melting at the lipid acyl chain and water position.
Data are normalized to monitor, detector efficiency normalized to
the scattering signal of a 2~mm thick Vanadium plate oriented at
135$^{\circ}$ with respect to the incoming beam. Please note that
no absorption correction has been done. For geometrical reasons
the diffraction detectors have a lower maximum $q_{||}$ than the
backscattering detectors.}\label{chainpeak}
\end{minipage}
\end{figure*}
Inelastic scattering techniques give a wave vector resolved access
to dynamical properties, i.e., excitation frequencies or
relaxation rates can be quantified at different internal length
scales. This is important to associate the dynamics with
individual modes of motion. A unique advantage lies in the
simultaneous access to both structural and dynamical properties by
measuring the scattering S($\vec{Q}$,$\omega$) as a function of
wave vector transfer $\vec{Q}$ and energy transfer $\omega$. 
We hereby report on $\mu$eV energy resolved spectra in the model
membrane system DMPC-d54, achieved by the neutron backscattering
technique. By analyzing the respective $Q$ dependence, we
simultaneously probe low energetic density fluctuations of lipid
acyl chains and interstitial water molecules, i.e., the water
layer in between the stacked membranes.

The spectrum of fluctuations in biomimetic and biological
membranes covers a large range of time and length scales
\cite{Koenig:1992,Koenig:1994,Koenig:1995,Pfeiffer:1989,Pfeiffer:1993,Lindahl:2000,Lipowsky:1995,Bayerl:2000,Salditt:2000},
ranging from the long wavelength undulation and bending modes of
the bilayer with typical relaxation times of nanoseconds and
lateral length scales of several hundred lipid molecules to the
short wavelength density fluctuations in the picosecond range on
nearest neighbor distances of lipid molecules. Local dynamics in
lipid bilayers, i.e., dynamics of individual lipid molecules as
vibration, rotation, libration (hindered rotation) and diffusion,
has been investigated by, e.g., incoherent neutron scattering
\cite{Koenig:1992,Koenig:1994,Koenig:1995,Pfeiffer:1989,Pfeiffer:1993}
and nuclear magnetic resonance \cite{Nevzorov:1997,Bloom:1995} to
determine the short wavelength translational and rotational
diffusion constant. Collective undulation modes have been
investigated using neutron spin-echo spectrometers
\cite{Pfeiffer:1989,Pfeiffer:1993,Takeda:1999,RheinstaedterPRL:2006}
and dynamical light scattering
\cite{Hirn:1998,Hirn:1999,Hildenbrand:2005}. Only recently, the
first inelastic scattering experiments in phospholipid bilayers to
determine collective motions of the lipid acyl chains and in
particular the short wavelength dispersion relation have been
performed using inelastic x-ray \cite{Chen:2001} and neutron
\cite{RheinstaedterPRL:2004} scattering techniques. While here
fast propagating sound modes in the picosecond time range have
been quantified, the present paper deals with slow nanosecond
relaxation times on length scales of nearest neighbor distances of
phospholipid acyl chains and water molecules, i.e., the slow
dynamics of melting (diffusion) and collective movements (most
likely undulations) of the lipid and water backbone. We have
selected the neutron backscattering technique for this study since
the dynamical modes at high $q_{||}$ (the in plane component of
the scattering vector $\vec{Q}$) are too fast to be accessed by
x-ray photon correlation spectroscopy (XPCS) and the lateral
length scales are too small to be resolved by dynamic light
scattering (DLS) or neutron spin-echo technique (NSE). Note that
the feasibility of wave vector resolved backscattering experiments
in oriented membrane systems to study freezing of lipid acyl
chains and water molecules has been established only recently
\cite{RheinstaedterPRE:2005}. In this work we present the
unprecedented determination of relaxation rates from Q-resolved
backscattering QENS data in phospholipid bilayers, combined with
in-situ diffraction to unambiguously assign the relaxation times
to the respective phases of the model membranes. These
measurements are carried out in aligned phases to preserve the
unique identification of modes on the basis of the parallel and
perpendicular components $q_{||}$, the lateral momentum transfer
in the plane of the bilayers, and $q_z$, the reflectivity, of the
scattering vector $\vec{Q}$.

\section{Experimental\label{Experimental}}
In the case of single membranes the inelastic neutron scattering
signal is by far not sufficient for a quantitative study of the
inelastic scattering. Partially (acyl chain) deuterated DMPC-d54
(1,2-dimyristoyl-sn-glycero-3-phoshatidylcholine) was obtained
from Avanti Polar Lipids. Highly oriented multi lamellar membrane
stacks of several thousands of lipid bilayers were prepared by
spreading lipid solution of typically 25mg/ml lipid in
trifluoroethylene/chloroform (1:1) on 2'' silicon wafers, followed
by subsequent drying in vacuum and hydration from D$_2$O vapor
\cite{Muenster:1999}, resulting in a structure of smectic A
symmetry. Twenty such wafers separated by small air gaps were
combined and aligned with respect to each other to create a
''sandwich sample'' consisting of several thousands of highly
oriented lipid bilayers (total mosaicity about 0.5$^{\circ}$),
with a total mass of about 400~mg of deuterated DMPC. The sample
was mounted in a hermetically sealed aluminium container within a
cryostat and hydrated from D$_2$O vapor. Saturation of the vapor
in the voids around the lipids was assured by placing a piece of
pulp soaked in D$_2$O within the sealed sample container. The pulp
was shielded by Cadmium to exclude any parasitic contribution to
the scattering. The hydration was not controlled but we allowed
the sample to equilibrate for 10~h at room temperature before the
measurements. The large beam divergence as compared to
reflectometers did not allow to determine the corresponding $d_z$
spacing simultaneously with sufficient accuracy to, e.g.,
determine the swelling state of the membrane stack. We note that
the absolute level of hydration is therefore not precisely known
but the temperature of the main transition agrees quite well with
literature values.

The experiment was carried out at the cold neutron backscattering
spectrometer IN16 \cite{Frick:2001} at the Institut Laue-Langevin
(ILL) in its standard setup with Si(111) monochromator and
analyzer crystals corresponding to an incident and analyzed
neutron energy of 2.08~meV ($\lambda$=6.27~\AA) resulting in a
high resolution in energy transfer of about 0.9~$\mu$eV~FWHM. An
energy transfer of $-15\mu$eV$<E<+15\mu$eV can be scanned by
varying the incident energy by Doppler-shifting the incident
neutron energy through an adequate movement of the monochromator
crystal. 20 detectors in exact backscattering with respect to the
analyzer crystals covering an angular width of 6.5$^{\circ}$ each
have been used. Note that the maximum of the static {\em bulk}
water correlation peak at about $q_{||}\approx$2~\AA$^{-1}$ is not
accessible because of simple geometrical reasons. The set-up is
therefore more sensitive to detect {\em amorphous} water with
slightly larger nearest neighbor distances, as it is expected to
occur in the water layers of the stacked bilayers. A separate line
of 160 energy integrating diffraction detectors -- normally
mounted on a parallel circle inclined by 24.4$^{\circ}$ below the
scattering plane -- with an angular width of 1$^{\circ}$ each --
allows to simultaneously detect structural changes with a much
higher $q_{||}$ resolution.
\begin{figure*} \centering
\resizebox{0.545\textwidth}{!}{\rotatebox{0}{\includegraphics{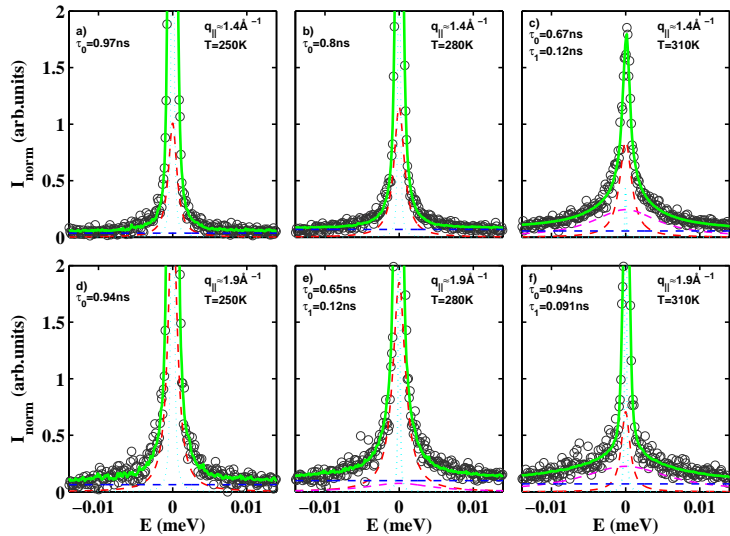}}}
\begin{minipage}[b]{8cm}
\centering \caption[]{(Color online). Inelastic scans at the lipid
chain ($q_{||}\approx$1.4~\AA$^{-1}$) and water
($q_{||}\approx$1.9~\AA$^{-1}$) position for temperatures T=250~K,
280~K and 310~K.  Detectors have been grouped to increase counting
statistics. The inter acyl chain correlation peak is measured by
detectors 9-13, which cover a $q_{||}$ range of
1.20~\AA$^{-1}$$<q_{||}$$<$1.60~\AA$^{-1}$. The water contribution
is detected by detectors 17-20 covering
1.78~\AA$^{-1}$$<q_{||}$$<$1.94~\AA$^{-1}$. As a model, up to two
Lorentzian peak profiles (dashed lines) to describe the quasi
elastic broadening and a Dirac function (dotted line) to describe
the elastic intensity were assumed. This model was convoluted with
the measured resolution function obtained from a Vanadium
standard. A flat background (dashed line) was subsequently added
and the result was fitted (fit result: solid line) to the data
(circles). A flat background may arise from fast processes far
beyond the accessible energy window of the spectrometer. The thus
obtained relaxation times are given in the figures. The
calibration error of the energy scale is approx.\@
3\%.}\label{inelastics}
\end{minipage}
\end{figure*}
\begin{figure*} \centering
\resizebox{0.545\textwidth}{!}{\rotatebox{0}{\includegraphics{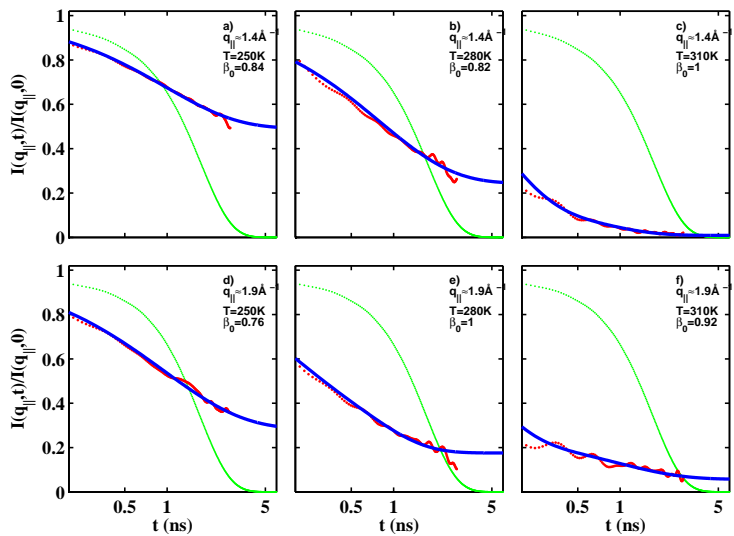}}}
\begin{minipage}[b]{8cm}
\centering \caption[]{(Color online). Corresponding Fourier
transformed energy space data (power spectra) from
Fig.~\ref{inelastics} (points), divided by an analytical
description of the Vanadium resolution. The thus obtained
intermediate scattering function $I(q_{||},t)/I(q_{||},0)$ is
shown together with the fits (blue solid line) of two exponential
decays $I(q_{||},t)/I(q_{||},0)=(A_0-A_1)\exp[-\left( t/\tau_0
(q_{||})\right)^{\beta0}]+y_1+(A_1-y_1)\exp[-\left( t/\tau_1
(q_{||})\right)^{\beta1}]$. The values for $\tau_{0,1}$ have been
fixed from the energy scans and only the exponents $\beta_{0,1}$
and the amplitudes $A_{0,1}$ have been fitted.  The thus obtained
values for $\beta_0$ are given in the figures. The resolution is
also plotted (green line). While the data at 250~K are well
described by one exponential decay, two decays are needed at
higher temperatures.}\label{timespace}
\end{minipage}
\end{figure*}

\section{Results from Neutron Diffraction\label{Diffraction}}
We thus simultaneously performed two types of diffraction
measurements, namely energy-resolved and energy-integrated
measurements.
Note that the additional
diffraction detectors 
very well integrate over low lying excitations and relaxations.
Figure~\ref{chainpeak} (a) shows an energy integrated diffraction
pattern for 1.1~\AA$^{-1}$$<q_{||}$$<$1.85~\AA$^{-1}$. From the
temperature dependence of the inter-acyl chain correlation peak at
$q_{||}\approx$1.4~\AA$^{-1}$ we assign the transition
temperatures of gel to ripple phase (L$\beta$-P$\beta'$) to 285~K
and the temperature of the main transition (P$\beta'$-L$\alpha$)
to T=295~K. T$_{fw}$=271~K marks the freezing temperature of the
amorphous water layer in between the stacked membranes
\cite{RheinstaedterPRE:2005}. The correlation peak gradually
shifts to smaller $q_{||}$ values below T$_{fw}$ until about
264~K. In the range 264~K$<$T$<$271~K water migrates out the
bilayer stacks to freeze as bulk water, as discussed in detail in
Ref.~[\onlinecite{RheinstaedterPRE:2005}]. The energy resolved
pattern in Fig.~\ref{chainpeak} (b) sheds light on the melting
process of lipid acyl chains and water and perfectly reproduces
the results previously reported in
[\onlinecite{RheinstaedterPRE:2005}] (the difference being the
larger number of detectors). Only structures which are static with
respect to the energy resolution of 0.9~$\mu$eV, which correspond
to motions slower than approx.\@ 4~ns, are detected as elastic
when using the energy resolving backscattering detectors. Changes
in this intensity may arise from either structural or dynamical
changes, i.e., shifts of correlation peaks or freezing or melting
of dynamical modes. Elastic intensity at the water detectors rises
at T$<$271~K; at the acyl chain position there is an elastic
contribution below 295~K, the temperature of the main transition.
A further increase occurs when entering the gel (L$\beta$) phase
indicating the high degree of dynamics in the ripple (P$\beta'$)
phase. Note that there is very little elastic contribution in the
fluid phases pointing out that the intensity usually measured in
diffraction experiments stems predominantly from low lying fluid
excitations and quasi elastic scattering.

\section{Inelastic Scattering\label{Inelastic}}
\begin{figure} \centering
\resizebox{1.00\columnwidth}{!}{\rotatebox{0}{\includegraphics{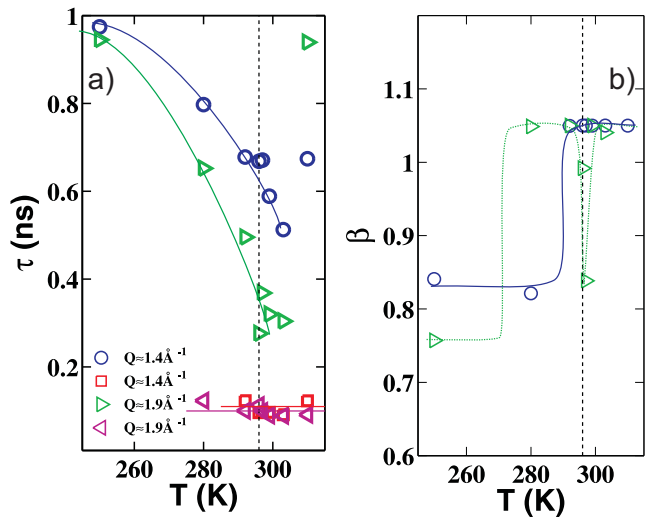}}}
 \centering \caption[]{(Color online).
(a) Relaxation times at the lipid chain and the water position for
all measured temperatures as determined from fits to the quasi
elastic data in Fig.~\ref{inelastics}. (b) Exponents of the
exponential decay as determined from fits of the intermediate
scattering function from Fig.~\ref{timespace}. Solid lines are
guides to the eye.} \label{tau}
\end{figure}

Based on the diffraction patterns we have recorded inelastic scans
at temperatures of T=250~K, 280~K, 292~K and at 310~K, far in the
fluid L$\alpha$ phase of the model membranes. Additional scans in
the regime of critical swelling (T=296, 297, 299 and 303~K) have
also been taken. The results for the acyl chain
($q_{||}\approx$1.4~\AA$^{-1}$) and the water position
($q_{||}\approx$1.9~\AA$^{-1}$) are depicted in
Fig.~\ref{inelastics}. At 250~K, both $q_{||}$ positions are well
fitted by a single Lorentzian, i.e., a single relaxation process
with an energy width (FWHM) $\Delta\omega$ and corresponding
relaxation time $\tau=2\pi/\Delta\omega$. For temperatures above
280~K for the water and above 292~K for the acyl chains two
Lorentzians are needed to describe the data. Although the
Lorentzians describe the energy spectra, small deviations from the
fits are visible pointing to deviations from single exponential
relaxations. A detailed analysis in energy space is difficult and
would mean to add more Lorentzian or mix Lorentzian and Gaussian
peak profiles with the risk of being physically meaningless. To
further characterize the relaxation processes the data have been
Fourier transformed (using the Matlab FFT algorithm) to determine
the exponent $\beta$ of the corresponding exponential decays in
time domain, $\exp{[-(t/\tau)^{\beta}]}$, from fits to the
intermediate scattering function $I(q_{||},t)$, as shown in
Fig.~\ref{timespace}. With increasing temperature there is a
relaxation step moving into the time window at the two $q_{||}$
positions, respectively. Because of the limited dynamical range of
15~$\mu$eV, the shortest time accessible is about 0.2~ns and the
faster process is already partially out of the accessible time
window, i.e., the quasielastic broadening is slightly broader than
the energy window. No exponent could therefore be reliably
determined for the fast process. The resulting relaxation times
$\tau_{0,1}$ and the corresponding exponent of the slow process,
$\beta_{0}$, are depicted in Fig.~\ref{tau} (a) and (b) for all
measured temperatures. To gain reliable fitting parameters for
$\tau$ and $\beta$, the relaxation times $\tau_{0,1}$ have been
determined from the energy spectra $S(q_{||},\omega)$ in
Fig.~\ref{inelastics} because this gives the highest accuracy and
stability. The $\tau$ values have then been fixed for the fits of
the exponential decay in the time domain. $\beta$ is determined
from the fits to $I(q_{||},t)$ in Fig.~\ref{timespace}.
The two processes at the two $q_{||}$ positions show distinctly
different relaxation times and temperature dependencies. While the
fast process has constant times of about 0.1~ns, the slow process
starts at about 1~ns at 250~K and becomes faster as approaching
the main phase transition at 295~K. Note that because of the
limited dynamical range the accuracy in determining the fast
relaxation time $\tau_1$ is also limited and we can not exclude a
temperature dependence of the fast process. While for the fast
branches no exponent can be determined for the same reason, the
slow processes show a stretched exponential character at low
temperatures and turn to single exponential when the corresponding
correlations melt at the lipid (main transition) and water melting
temperature, respectively.
\section{Discussion\label{Discussion}}
From the inelastic neutron data we quantify two relaxation times.
We argue that the corresponding processes stem from collective
motions of the lipid acyl chains and interstitial water molecules,
respectively, rather than from local dynamics. By selective
deuteration of the chains and hydration from D$_2$O vapor,
coherent motions are strongly enhanced over other contributions to
the inelastic scattering cross section \footnote{Neutron
scattering data may include contributions from coherent and
incoherent scattering, i.e., from pair- and autocorrelated
scattering (collective and local or diffusive modes). While in
protonated samples the {\em incoherent} scattering is usually
dominant and the time-autocorrelation function of individual
scatterers is accessible in neutron scattering experiments,
(partial) deuteration emphasizes the {\em coherent} scattering and
gives access to collective motions by probing the pair correlation
function.}. The fast process $\tau_1$ disappears below the
freezing temperature of the lipids (P$\beta'$ phase) respective
water molecules (T$<$T$_{fw}$=271~K) and might therefore be due to
(collective) translational diffusion of lipids and water. The slow
process with relaxation times $\tau_0$ is measured down to the
lowest temperatures of T=250~K and most likely probes collective
dynamics, i.e., neighboring lipid and water molecules
participating in slow mesoscopic dynamics. However, to further
characterize this process, $q_{||}$ dependent data are needed to
determine the corresponding dispersion relation and compare to
theoretical models. It is speculated in the literature that the
minimum in the dispersion relation of the collective fast
propagating modes at the nearest neighbor distances of lipid acyl
chains (as determined in Refs.~[\onlinecite{Chen:2001}] and
[\onlinecite{RheinstaedterPRL:2004}]) is related to transport
phenomena within and across the bilayers. The relaxation found in
the present paper on the same length but distinctly different time
scale might be relevant to establish such a model of {\em phonon
assisted diffusion} in membranes, which would be of particular
interest in membrane biophysics and biotechnology applications.

The slow relaxations at the acyl chain and the water position both
show a decrease of relaxation times towards the main transition of
the phospholipids. 
The relaxation time of the fast branch with $\tau\approx$0.1~ns
that we attribute to translational diffusion is temperature
independent within the experimental accuracy, which is slightly
limited by the maximum achievable energy transfer. Note that the
diffusion time measured here can be interpreted as time for
hopping processes to nearest neighbor sites. Such a process needs
a free neighboring site but does not necessarily give information
about the mobility of these {\em holes} (free surface or volume)
what would be important for macroscopic diffusion. The
corresponding time scales and diffusion constants might therefore
be distinctly different. From measurements with a higher dynamical
range, the energy barrier of this excitation could probably be
determined from the temperature dependent hopping times using
Arrhenius or Vogel-Fulcher laws.

The $\beta$ exponents in Fig.~\ref{tau} (b) start at values
$\beta<1$ at low temperatures. Structural inhomogeneities and
heterogeneous interactions obviously lead to a local relaxation
dynamics and to stretched ($\beta<1$) exponentials. When entering
the fluid phase at T$_m$, the lipid relaxations
(q$_{||}\approx1.4$~\AA$^{-1}$) turn into single exponential, as
can be expected for a diffusive, fluid like motion of the
particles. At the water position (q$_{||}\approx1.9$~\AA$^{-1}$),
the transition from stretched to single exponential occurs between
250~K and 280~K, most likely at the water freezing or melting
temperature T$_{fw}$=271~K, as speculated and indicated in the
Figure. A striking feature is that in the range of critical
swelling, the water relaxations become again stretched what might
be related to the expansion of the water layer leading to the well
known anomalous swelling of phospholipid bilayers close to the
main transition \cite{Pabst:2003}.

\section{Conclusion\label{Conclusion}}
In conclusion, using the neutron backscattering technique, we have
determined $q_{||}$-dependent relaxation times at different
temperatures in the model membrane system DMPC that we attribute
to lipid acyl chains and interstitial water molecules. In-situ
diffraction allowed to clearly assign the temperatures to the
different phases of the bilayers. A combined data analysis in the
energy and time domain allowed to quantify relaxation times of two
processes and determine the exponents $\beta$ of the exponential
decays. The fast process is attributed to diffusion while the slow
process likely stems from collective dynamics.
We find stretched exponential relaxations at low
temperatures. The relaxations turn into single exponential above
the corresponding melting temperatures of lipid and water
molecules.

Future experiments, which will include a selective deuteration of
the acyl-chains and the membrane water, respectively, will allow
to mask different types of mobility and hopefully deduce complete
dispersion relations of the different molecular components in the
different phases of the phospholipid bilayers. Compared to NSE and
XPCS, large Q-values can easily be obtained. The current drawback
of the technique is the limited dynamical range on high-flux
spectrometers which hopefully will be overcome in next generation
neutron sources and backscattering spectrometers. Our study also
points to the wealth of new information that can be explored with
the membrane backscattering technique.

{\bf Acknowledgement:} We acknowledge T.~Gronemann (Institut f\"ur
R\"ontgenphysik, G\"ottingen) for help with the sample preparation
and M.~Elender (ILL) for technical and engineering support and the
ILL for the allocation of beam time.

\bibliography{./Membranes_Rheinstaedter_09072006}

\begin{thebibliography}{22}
\expandafter\ifx\csname natexlab\endcsname\relax\def\natexlab#1{#1}\fi
\expandafter\ifx\csname bibnamefont\endcsname\relax
  \def\bibnamefont#1{#1}\fi
\expandafter\ifx\csname bibfnamefont\endcsname\relax
  \def\bibfnamefont#1{#1}\fi
\expandafter\ifx\csname citenamefont\endcsname\relax
  \def\citenamefont#1{#1}\fi
\expandafter\ifx\csname url\endcsname\relax
  \def\url#1{\texttt{#1}}\fi
\expandafter\ifx\csname urlprefix\endcsname\relax\def\urlprefix{URL }\fi
\providecommand{\bibinfo}[2]{#2}
\providecommand{\eprint}[2][]{\url{#2}}

\bibitem[{\citenamefont{K\"onig et~al.}(1992)\citenamefont{K\"onig, Pfeiffer,
  Bayerl, Richter, and Sackmann}}]{Koenig:1992}
\bibinfo{author}{\bibfnamefont{S.}~\bibnamefont{K\"onig}},
  \bibinfo{author}{\bibfnamefont{W.}~\bibnamefont{Pfeiffer}},
  \bibinfo{author}{\bibfnamefont{T.}~\bibnamefont{Bayerl}},
  \bibinfo{author}{\bibfnamefont{D.}~\bibnamefont{Richter}}, \bibnamefont{and}
  \bibinfo{author}{\bibfnamefont{E.}~\bibnamefont{Sackmann}},
  \bibinfo{journal}{J. Phys. II France} \textbf{\bibinfo{volume}{2}},
  \bibinfo{pages}{1589} (\bibinfo{year}{1992}).

\bibitem[{\citenamefont{K\"onig et~al.}(1994)\citenamefont{K\"onig, Sackmann,
  Richter, Zorn, Carlile, and Bayerl}}]{Koenig:1994}
\bibinfo{author}{\bibfnamefont{S.}~\bibnamefont{K\"onig}},
  \bibinfo{author}{\bibfnamefont{E.}~\bibnamefont{Sackmann}},
  \bibinfo{author}{\bibfnamefont{D.}~\bibnamefont{Richter}},
  \bibinfo{author}{\bibfnamefont{R.}~\bibnamefont{Zorn}},
  \bibinfo{author}{\bibfnamefont{C.}~\bibnamefont{Carlile}}, \bibnamefont{and}
  \bibinfo{author}{\bibfnamefont{T.}~\bibnamefont{Bayerl}},
  \bibinfo{journal}{J. Chem. Phys.} \textbf{\bibinfo{volume}{100}},
  \bibinfo{pages}{3307} (\bibinfo{year}{1994}).

\bibitem[{\citenamefont{K\"onig et~al.}(1995)\citenamefont{K\"onig, Bayerl,
  Coddens, Richter, and Sackmann}}]{Koenig:1995}
\bibinfo{author}{\bibfnamefont{S.}~\bibnamefont{K\"onig}},
  \bibinfo{author}{\bibfnamefont{T.}~\bibnamefont{Bayerl}},
  \bibinfo{author}{\bibfnamefont{G.}~\bibnamefont{Coddens}},
  \bibinfo{author}{\bibfnamefont{D.}~\bibnamefont{Richter}}, \bibnamefont{and}
  \bibinfo{author}{\bibfnamefont{E.}~\bibnamefont{Sackmann}},
  \bibinfo{journal}{Biophys. J.} \textbf{\bibinfo{volume}{68}},
  \bibinfo{pages}{1871} (\bibinfo{year}{1995}).

\bibitem[{\citenamefont{Pfeiffer et~al.}(1989)\citenamefont{Pfeiffer, Henkel,
  Sackmann, and Knorr}}]{Pfeiffer:1989}
\bibinfo{author}{\bibfnamefont{W.}~\bibnamefont{Pfeiffer}},
  \bibinfo{author}{\bibfnamefont{T.}~\bibnamefont{Henkel}},
  \bibinfo{author}{\bibfnamefont{E.}~\bibnamefont{Sackmann}}, \bibnamefont{and}
  \bibinfo{author}{\bibfnamefont{W.}~\bibnamefont{Knorr}},
  \bibinfo{journal}{Europhys. Lett.} \textbf{\bibinfo{volume}{8}},
  \bibinfo{pages}{201} (\bibinfo{year}{1989}).

\bibitem[{\citenamefont{Pfeiffer et~al.}(1993)\citenamefont{Pfeiffer, K\"onig,
  Legrand, Bayerl, Richter, and Sackmann}}]{Pfeiffer:1993}
\bibinfo{author}{\bibfnamefont{W.}~\bibnamefont{Pfeiffer}},
  \bibinfo{author}{\bibfnamefont{S.}~\bibnamefont{K\"onig}},
  \bibinfo{author}{\bibfnamefont{J.}~\bibnamefont{Legrand}},
  \bibinfo{author}{\bibfnamefont{T.}~\bibnamefont{Bayerl}},
  \bibinfo{author}{\bibfnamefont{D.}~\bibnamefont{Richter}}, \bibnamefont{and}
  \bibinfo{author}{\bibfnamefont{E.}~\bibnamefont{Sackmann}},
  \bibinfo{journal}{Europhys. Lett.} \textbf{\bibinfo{volume}{23}},
  \bibinfo{pages}{457} (\bibinfo{year}{1993}).

\bibitem[{\citenamefont{Lindahl and Edholm}(2000)}]{Lindahl:2000}
\bibinfo{author}{\bibfnamefont{E.}~\bibnamefont{Lindahl}} \bibnamefont{and}
  \bibinfo{author}{\bibfnamefont{O.}~\bibnamefont{Edholm}},
  \bibinfo{journal}{Biophys. J.} \textbf{\bibinfo{volume}{79}},
  \bibinfo{pages}{426} (\bibinfo{year}{2000}).

\bibitem[{\citenamefont{Lipowsky and Sackmann}(1995)}]{Lipowsky:1995}
\bibinfo{editor}{\bibfnamefont{R.}~\bibnamefont{Lipowsky}} \bibnamefont{and}
  \bibinfo{editor}{\bibfnamefont{E.}~\bibnamefont{Sackmann}}, eds.,
  \emph{\bibinfo{title}{Structure and Dynamics of Membranes}},
  vol.~\bibinfo{volume}{1} of \emph{\bibinfo{series}{Handbook of Biological
  Physics}} (\bibinfo{publisher}{Elsevier}, \bibinfo{address}{Amsterdam},
  \bibinfo{year}{1995}).

\bibitem[{\citenamefont{Bayerl}(2000)}]{Bayerl:2000}
\bibinfo{author}{\bibfnamefont{T.}~\bibnamefont{Bayerl}},
  \bibinfo{journal}{Curr. Opin. Colloid Interface Sci.}
  \textbf{\bibinfo{volume}{5}}, \bibinfo{pages}{232} (\bibinfo{year}{2000}).

\bibitem[{\citenamefont{Salditt}(2000)}]{Salditt:2000}
\bibinfo{author}{\bibfnamefont{T.}~\bibnamefont{Salditt}},
  \bibinfo{journal}{Curr. Opin. Colloid Interface Sci.}
  \textbf{\bibinfo{volume}{5}}, \bibinfo{pages}{19} (\bibinfo{year}{2000}).

\bibitem[{\citenamefont{Nevzorov and Brown}(1997)}]{Nevzorov:1997}
\bibinfo{author}{\bibfnamefont{A.}~\bibnamefont{Nevzorov}} \bibnamefont{and}
  \bibinfo{author}{\bibfnamefont{M.}~\bibnamefont{Brown}}, \bibinfo{journal}{J.
  Chem. Phys.} \textbf{\bibinfo{volume}{107}}, \bibinfo{pages}{10288}
  (\bibinfo{year}{1997}).

\bibitem[{\citenamefont{Bloom and Bayerl}(1995)}]{Bloom:1995}
\bibinfo{author}{\bibfnamefont{M.}~\bibnamefont{Bloom}} \bibnamefont{and}
  \bibinfo{author}{\bibfnamefont{T.}~\bibnamefont{Bayerl}},
  \bibinfo{journal}{Can. J. Phys.} \textbf{\bibinfo{volume}{73}},
  \bibinfo{pages}{687} (\bibinfo{year}{1995}).

\bibitem[{\citenamefont{Takeda et~al.}(1999)\citenamefont{Takeda, Kawabata,
  Seto, Komura, Gosh, Nagao, and Okuhara}}]{Takeda:1999}
\bibinfo{author}{\bibfnamefont{T.}~\bibnamefont{Takeda}},
  \bibinfo{author}{\bibfnamefont{Y.}~\bibnamefont{Kawabata}},
  \bibinfo{author}{\bibfnamefont{H.}~\bibnamefont{Seto}},
  \bibinfo{author}{\bibfnamefont{S.}~\bibnamefont{Komura}},
  \bibinfo{author}{\bibfnamefont{S.}~\bibnamefont{Gosh}},
  \bibinfo{author}{\bibfnamefont{M.}~\bibnamefont{Nagao}}, \bibnamefont{and}
  \bibinfo{author}{\bibfnamefont{D.}~\bibnamefont{Okuhara}},
  \bibinfo{journal}{J. Phys. Chem. Solids} \textbf{\bibinfo{volume}{60}},
  \bibinfo{pages}{1375} (\bibinfo{year}{1999}).

\bibitem[{\citenamefont{Rheinst\"adter
  et~al.}(2006)\citenamefont{Rheinst\"adter, H\"aussler, and
  Salditt}}]{RheinstaedterPRL:2006}
\bibinfo{author}{\bibfnamefont{M.~C.} \bibnamefont{Rheinst\"adter}},
  \bibinfo{author}{\bibfnamefont{W.}~\bibnamefont{H\"aussler}},
  \bibnamefont{and} \bibinfo{author}{\bibfnamefont{T.}~\bibnamefont{Salditt}},
  \bibinfo{journal}{accepted for publication}  (\bibinfo{year}{2006}),
  \bibinfo{note}{cond-mat/0606114}.

\bibitem[{\citenamefont{Hirn et~al.}(1998)\citenamefont{Hirn, Bayerl, R\"adler,
  and Sackmann}}]{Hirn:1998}
\bibinfo{author}{\bibfnamefont{R.}~\bibnamefont{Hirn}},
  \bibinfo{author}{\bibfnamefont{T.}~\bibnamefont{Bayerl}},
  \bibinfo{author}{\bibfnamefont{J.}~\bibnamefont{R\"adler}}, \bibnamefont{and}
  \bibinfo{author}{\bibfnamefont{E.}~\bibnamefont{Sackmann}},
  \bibinfo{journal}{Faraday Discuss.} \textbf{\bibinfo{volume}{111}},
  \bibinfo{pages}{17} (\bibinfo{year}{1998}).

\bibitem[{\citenamefont{Hirn and Bayerl}(1999)}]{Hirn:1999}
\bibinfo{author}{\bibfnamefont{R.~B.} \bibnamefont{Hirn}} \bibnamefont{and}
  \bibinfo{author}{\bibfnamefont{T.~M.} \bibnamefont{Bayerl}},
  \bibinfo{journal}{Phys. Rev. E} \textbf{\bibinfo{volume}{59}},
  \bibinfo{pages}{5987} (\bibinfo{year}{1999}).

\bibitem[{\citenamefont{Hildenbrand and Bayerl}(2005)}]{Hildenbrand:2005}
\bibinfo{author}{\bibfnamefont{M.~F.} \bibnamefont{Hildenbrand}}
  \bibnamefont{and} \bibinfo{author}{\bibfnamefont{T.~M.}
  \bibnamefont{Bayerl}}, \bibinfo{journal}{Biophys. J.}
  \textbf{\bibinfo{volume}{88}}, \bibinfo{pages}{3360} (\bibinfo{year}{2005}).

\bibitem[{\citenamefont{Chen et~al.}(2001)\citenamefont{Chen, Liao, Huang,
  Weiss, Bellisent-Funel, and Sette}}]{Chen:2001}
\bibinfo{author}{\bibfnamefont{S.}~\bibnamefont{Chen}},
  \bibinfo{author}{\bibfnamefont{C.}~\bibnamefont{Liao}},
  \bibinfo{author}{\bibfnamefont{H.}~\bibnamefont{Huang}},
  \bibinfo{author}{\bibfnamefont{T.}~\bibnamefont{Weiss}},
  \bibinfo{author}{\bibfnamefont{M.}~\bibnamefont{Bellisent-Funel}},
  \bibnamefont{and} \bibinfo{author}{\bibfnamefont{F.}~\bibnamefont{Sette}},
  \bibinfo{journal}{Phys. Rev. Lett.} \textbf{\bibinfo{volume}{86}},
  \bibinfo{pages}{740} (\bibinfo{year}{2001}).

\bibitem[{\citenamefont{Rheinst\"adter
  et~al.}(2004)\citenamefont{Rheinst\"adter, Ollinger, Fragneto, Demmel, and
  Salditt}}]{RheinstaedterPRL:2004}
\bibinfo{author}{\bibfnamefont{M.~C.} \bibnamefont{Rheinst\"adter}},
  \bibinfo{author}{\bibfnamefont{C.}~\bibnamefont{Ollinger}},
  \bibinfo{author}{\bibfnamefont{G.}~\bibnamefont{Fragneto}},
  \bibinfo{author}{\bibfnamefont{F.}~\bibnamefont{Demmel}}, \bibnamefont{and}
  \bibinfo{author}{\bibfnamefont{T.}~\bibnamefont{Salditt}},
  \bibinfo{journal}{Phys. Rev. Lett.} \textbf{\bibinfo{volume}{93}},
  \bibinfo{pages}{108107} (\bibinfo{year}{2004}).

\bibitem[{\citenamefont{Rheinst\"adter
  et~al.}(2005)\citenamefont{Rheinst\"adter, Seydel, Demmel, and
  Salditt}}]{RheinstaedterPRE:2005}
\bibinfo{author}{\bibfnamefont{M.~C.} \bibnamefont{Rheinst\"adter}},
  \bibinfo{author}{\bibfnamefont{T.}~\bibnamefont{Seydel}},
  \bibinfo{author}{\bibfnamefont{F.}~\bibnamefont{Demmel}}, \bibnamefont{and}
  \bibinfo{author}{\bibfnamefont{T.}~\bibnamefont{Salditt}},
  \bibinfo{journal}{Phys. Rev. E} \textbf{\bibinfo{volume}{71}},
  \bibinfo{pages}{061908} (\bibinfo{year}{2005}).

\bibitem[{\citenamefont{M\"unster et~al.}(1999)\citenamefont{M\"unster,
  Salditt, Vogel, Siebrecht, and Peisl}}]{Muenster:1999}
\bibinfo{author}{\bibfnamefont{C.}~\bibnamefont{M\"unster}},
  \bibinfo{author}{\bibfnamefont{T.}~\bibnamefont{Salditt}},
  \bibinfo{author}{\bibfnamefont{M.}~\bibnamefont{Vogel}},
  \bibinfo{author}{\bibfnamefont{R.}~\bibnamefont{Siebrecht}},
  \bibnamefont{and} \bibinfo{author}{\bibfnamefont{J.}~\bibnamefont{Peisl}},
  \bibinfo{journal}{Europhys. Lett.} \textbf{\bibinfo{volume}{46}},
  \bibinfo{pages}{486} (\bibinfo{year}{1999}).

\bibitem[{\citenamefont{Frick and Gonzalez}(2001)}]{Frick:2001}
\bibinfo{author}{\bibfnamefont{B.}~\bibnamefont{Frick}} \bibnamefont{and}
  \bibinfo{author}{\bibfnamefont{M.}~\bibnamefont{Gonzalez}},
  \bibinfo{journal}{Physica B} \textbf{\bibinfo{volume}{301}},
  \bibinfo{pages}{8} (\bibinfo{year}{2001}).

\bibitem[{\citenamefont{Pabst et~al.}(2003)\citenamefont{Pabst, Katsaras,
  Raghunathan, and Rappolt}}]{Pabst:2003}
\bibinfo{author}{\bibfnamefont{G.}~\bibnamefont{Pabst}},
  \bibinfo{author}{\bibfnamefont{J.}~\bibnamefont{Katsaras}},
  \bibinfo{author}{\bibfnamefont{V.~A.} \bibnamefont{Raghunathan}},
  \bibnamefont{and} \bibinfo{author}{\bibfnamefont{M.}~\bibnamefont{Rappolt}},
  \bibinfo{journal}{Langmuir} \textbf{\bibinfo{volume}{19}},
  \bibinfo{pages}{1716} (\bibinfo{year}{2003}).

\end{thebibliography}

\end{document}